\documentclass[journal=jpccck,manuscript=article]{achemso}
\usepackage[version=3]{mhchem} 

\usepackage{amsmath,amssymb}
\usepackage{amsfonts}
\usepackage{bm}
\usepackage{stackengine}
\stackMath
\usepackage{natbib}
\usepackage{graphicx}
\usepackage{array}
\usepackage{dcolumn}
\usepackage{amssymb}
\usepackage{multirow}
\usepackage{mathtools}
\newcolumntype{z}[1]{D{.}{.}{#1}}
\usepackage{xcolor}
\usepackage{hyperref}
\usepackage{float}
\usepackage{mciteplus}
\mciteErrorOnUnknownfalse 
\usepackage{tabularx}
\usepackage{rotating}
\usepackage{placeins}
\usepackage{romannum}
\hypersetup{
    colorlinks=true,
    linkcolor=black,
    citecolor=black,
    filecolor=black,
    urlcolor=black,
}
    
\renewcommand*{\equationautorefname}{Eq.}
\renewcommand*{\figureautorefname}{Fig.}
\usepackage{physics}
\title{Entangled Photon Resonance Energy Transfer in Arbitrary Media}
\date{\today}
\author{K.~Nasiri~Avanaki}
\affiliation{ Department of Chemistry, Northwestern University, 2145 Sheridan Road, Evanston IL 60208-3113,USA}
\alsoaffiliation{ Department of Chemistry and Chemical Biology, Harvard University, 12 Oxford Street, Cambridge MA 02138, USA}
\author{George C. Schatz}
\affiliation{ Department of Chemistry, Northwestern University, 2145 Sheridan Road, Evanston IL 60208-3113,USA}
\email{g-schatz@northwestern.edu}

\SectionNumbersOn

\begin{document}
\pagenumbering{arabic}
\begin{abstract}
\end{abstract}
\maketitle
Inspired by the unique nonclassical character of two-photon interactions induced by entangled photons, we develop a new  comprehensive F\"orster-type
formulation for entangled two-photon resonance energy transfer (E2P-RET) mediated by inhomogeneous, dispersive and absorptive media with any space-dependent and frequency-dependent dielectric function and with any size of donor/acceptor. In our theoretical framework, two uncoupled particles are jointly excited by the temporally entangled field associated with two virtual photons that are produced by three-level radiative cascade decay in a donor particle.
The temporal entanglement leads to frequency anticorrelation in the virtual photon's field, and vanishing of one of the time-ordered excitation pathways. The underlying mechanism leads to more than three orders of magnitude enhancement in the E2P-RET rate compared with the uncorrelated photon case. With the power of our new formulation, we propose a way to characterize E2P-RET through an effective rate coefficient $K_{E2P}$, introduced here. This coefficient shows how energy transfer can be enhanced or suppressed depending on rate parameters in the radiative cascade, and by varying the donor-acceptor frequency differences.


\doublespacing
   \section{Introduction}
   
 Excitation energy transfer including radiative and
non-radiative mechanisms, is a universally important photophysical process in photoactive systems
defined as the relocation of electronic excitation energy from an optically excited donor to a nearby acceptor.  The originally formulated F\"{o}rster theory \cite{Forster_1948} can describe RET in various problems and with some changes it can be utilized in more efficient hybrid systems  \cite{Gray_PRA2016,Gray_PRB2015,Ebbesen_Angewandte2017,EbbesenExp_Angewandte2016,Nanoshell_PRA2016} as well.

With the design and synthesis of multi-chromophore macromolecules, a new theoretical framework has been developed for the general case of twin-donor RET
\cite{Andrew4Center_2002} in the vicinity of an acceptor  \cite{AQED_theory_JPC2001,ThreeCenter_Andrews98,Multichromophore_Andrew_JCP2003,2Pfluorescence_AndrewsJCP98}, which is of interest for biomimetic energy conversion. These systems  capture optical radiation with high efficiency due to the large number of antenna chromophores and efficient mechanisms for channeling energy to an acceptor core\cite{OpticallyNonlinear_Andrew2004}.

In another direction, with advances in non-classical light sources \cite{Teich_PhysToday1990,Optical-Coherence_Book95,Generation_PRL2016,Energy-Time-Entangled_PRL2018,EPspectroscopy_JPB2017} and their application in exploring new phenomena in multiphoton processes, there has been a rebirth of interest and extensive attention in nonlinear laser spectroscopy involving entangled photons, perhaps most prominently in two-photon absorption/emission as fundamental components of non-classical light-matter interaction.
The features of quantum light open up a new era for discovery of valuable information on relaxation, transport pathways,  spectroscopy at extremely low input photon fluxes\cite{EP_Virtual-StatePRL1998,Entangled-biphoton_CPL2004}, entanglement-induced two-photon transparency
\cite{Entanglement-Induced_PRL1997} and entangled-photon
virtual-state spectroscopy\cite{EPspectroscopy_JPB2017,Goodson_JACKS2010,Optically_Excited_Goodson2013}. These fascinating developments cannot be retrieved from the linear response of the system interacting with the classical form of the light.
Today we have access to a variety of techniques for producing quantum light \cite{sphotons_APL2003,Squeezed_Opt2013}, entangled coherent states\cite{ECstates_PRA92}, and E2P states from spontaneous parametric down-conversion (SPDC)\cite{SPDC_PRL95,GeneratingE2P_PRL2002} widely used in quantum information, data encryption\cite{ECryptography_PRL2000,EntanglementOspin_Nat2010,EfficientQC_Nat2011}
and quantum communication \cite{UQteleportation_Sci2014,QTeleportation_Sci2009}.


In recent experiments \cite{Goodson_JACKS2010,Optically_Excited_Goodson2013,Entangled_Photon_Goodson2017} utilizing the SPDC technique, the phenomenon of entangled two-photon absorption (E2PA) interestingly showed linear rather than quadratic dependence of the absorption rate on excitation intensity which was dominant at low intensity. 
Indeed, the non-classical approaches involving two or more entangled photons  provide exceptional efficiency over conventional incoherent light sources. This motivates the present work. 

In this paper we seek to understand the underlying mechanism of RET for a system consisting of a single excited donor and a pair of uncoupled acceptors, with RET involving an entangled virtual pair of photons. The main goal here is to explore if the quantum state of light can give us better control/enhancement for the RET rate. This work thus bridges between the two fields of quantum optics and resonance energy migration in photoactive materials, and a goal of our analysis is to develop the theory of RET in arbitrary media and going beyond the electric dipole approximation. 

We assume that the source of entangled photons is a biexciton cascade \cite{Cascade_APL2007,Cascade_PRL2008,Optics_Book94} that takes place in a single particle quantum dot (QD). This provides us with a table-top source of “triggered” entangled virtual photon pairs, as it can produce no more than two photons per excitation cycle\cite{Cascade_APL2007}.
Indeed, using pulsed excitation, the two emitted  photons are
''clocked'' with one appearing shortly after the other.\cite{EP_Generation_2007,Cascade_Semicon_PRL2004,Cascade_Semicon_PRL2000,Cascade_Semicon_PRB2005}
\begin{figure*}[h]
\begin{center}
\includegraphics[width=.4\textwidth]{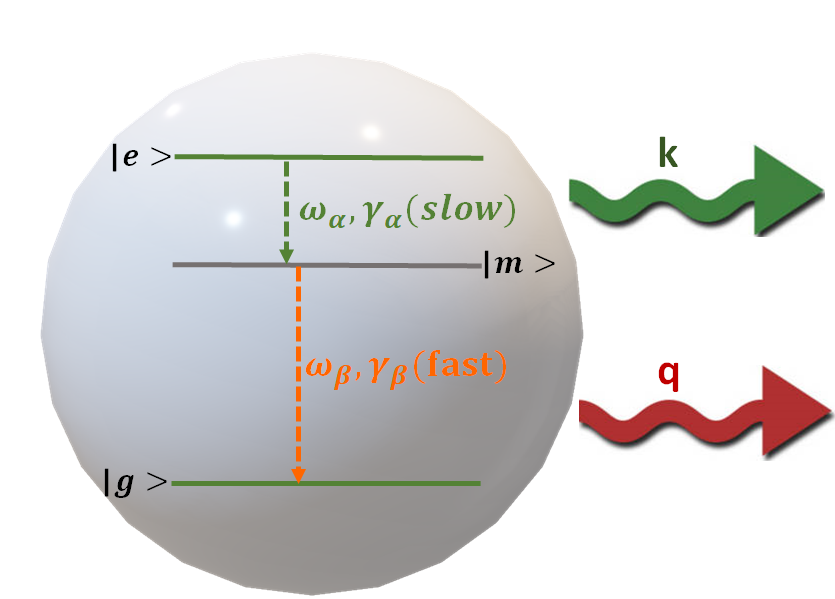}
\caption{The schematic generation of a two-photon state from the excited cascade of a three-level donor system. Initially the system is in excited state $\ket{e}$, and through the sequential emissions, a photon pair is generated with frequency anti-correlation due to energy conservation.}
\label{Fig:Decay}
\end{center}
\end{figure*}
And the entangled state in this case is both
correlated in time and anti-correlated in frequency.

In general, for any frequencies $\omega_{1}$ and $\omega_{2}$, if the
cross frequency correlation function satisfies $g^{(2)}_{\times}(r,\omega)=1$ \cite{Loudon_QTL}, the bipartite light beam is factorable; otherwise, the light beam has some frequency correlations between parts.
In a bipartite two-photon state with total energy of $\hbar( \omega_{1}+\omega_{2})=E_{2p}$, frequency anti-correlation means that if there is one photon in the $\omega_{1}$-frequency mode in the first partite, there is a higher probability to find the other photon in ($E_{2p}-\omega_{1}$)-frequency mode in the second partite. We employ this concept in our method  for the cascade emission from the three level QD source (see \autoref{Fig:Decay}) with a small width $\gamma_{\alpha}$ and a large width $\gamma_{\beta}$. 
The entangled state which is produced by this  emission process involves a single particle donor.  
\begin{figure*}[h]
\begin{center}
\includegraphics[width=0.9\textwidth]{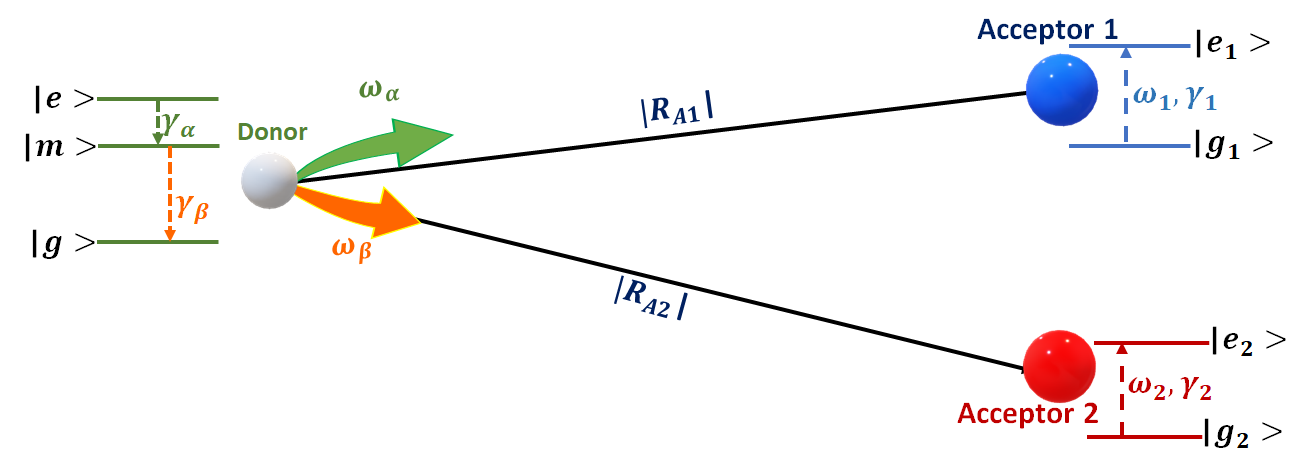}\\
\includegraphics[width=.6\textwidth]{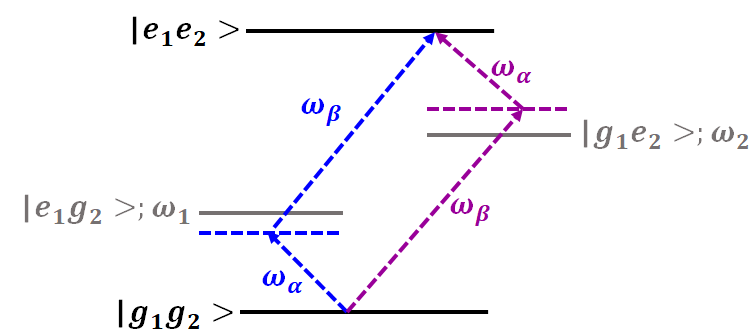}
\caption{Schematic of the two-acceptor interacting with E2P-model. There is no interaction between the two acceptors, and the  frequencies of the excited
states are $\omega_{1}$, $\omega_{2}$ respectively. The central frequencies of the two incident beams,
$\omega_{\alpha}$ and $\omega_{\beta}$, are far from resonance with the single-particle but their sum is almost equal to the
sum of the excitation energy of the acceptors i.e., $\delta=\omega_{\alpha}+\omega_{\beta}-\omega_{1}-\omega_{2}\approx 0 $.
}
\label{Fig:A1A2Source}
\end{center}
\end{figure*}

We assume that RET involves absorption of the two photons by a pair of uncoupled particles (see \autoref{Fig:A1A2Source}) taken to be  two-level systems (ground and excited states $\ket{g_{i}}$ and $\ket{e_{i}}$, $i =
1,2$). Generally the acceptors are not identical with corresponding excitation energies $\hbar \omega_{i}$, and spontaneous
emission rates $\gamma_{i}$. We assume
that the mean excitation time for each particle is much shorter than the
lifetimes of the two excited acceptors so that we can consider
that the two excited states have infinite lifetimes ($\gamma_{1,2}\approx 0$)
and maintain their excitation forever. 
Under the rotating-wave approximation
the Hamiltonian of the system can be written as
\begin{eqnarray}
  \begin{aligned}
  H&=H_{0}+V_{int}\\
  &=\hbar\omega_{1}b_{1}b^{\dagger}_{1}+\hbar\omega_{2}b_{2}b^{\dagger}_{2}+\hbar\sum_{\lambda=1,2}\sum_{l}\omega_{l} a^{\lambda}_{l}a^{\lambda\dagger}_{l}+V_{int}
  \label{Hamiltonian_2A}
  \end{aligned}
  \end{eqnarray}
where $b_{i}=\ket{g_{i}}\bra{e_{i}}$ and the annihilation operators $a_{l}$ are time-independent. Here $[a^{\lambda}_{l},a^{\lambda '\dagger}_{l'}]=\delta_{\lambda, \lambda'}\delta_{l,l'}$ and the interaction potential, $V_{int}$ is defined as
\begin{eqnarray}
  \begin{aligned}
  V_{int}(t)&=-\sum_{i}\boldsymbol{\mu}_{i}(b_{i}+b_{i}^{\dagger}).
  \left[
  \boldsymbol{E}_{i}^{(+)}(\bold{R}_{i},t)+\boldsymbol{E}_{i}^{(+)\dagger}(\bold{R}_{i},t)
  \right]
  \label{V_int1}
  \end{aligned}
  \end{eqnarray}
Here $\boldsymbol{\mu}_{i}$ is
the electric dipole transition of acceptor $i$ while $\boldsymbol{E}^{(+)}+\boldsymbol{E}^{(+)\dagger}$ is the emitted induced electric field generally written as a mode expansion \cite{ResonanceenergyAndrews2003,OpticallyNonlinear_Andrew2004}, arising from donor at the position of the acceptor $\bold{R}_{i}$ (where donor is placed at $\bold{r}_{D}$,  the acceptor $i$ is at $\bold{r}_{A_{i}}$, and $\bold{R}_{i}=\bold{r}_{A_{i}}-\bold{r}_{D}$),
\begin{eqnarray}
  \begin{aligned}
  \boldsymbol{E}^{(+)}(\bold{R}_{i},t)=i\sum_{\lambda=1,2}\sum_{l} \boldsymbol{\epsilon}_{l}^{(\lambda)}\hat{a}_{l}^{(\lambda)}e^{i(\bold{k}_{l}.\bold{R}_{i}-\omega_{l}t)}
  \label{V_int1}
  \end{aligned}
  \end{eqnarray}
In \autoref{V_int1}$,   \boldsymbol{\epsilon}^{(\lambda)}_{l}=\sqrt{\dfrac{\hbar \omega_{l}}{2 \epsilon_{0}V}}\boldsymbol{e}^{(\lambda)}_{l}$ and the two $\boldsymbol{e}^{(\lambda)}_{l}$ (polarization vectors) are conventional unit vectors for left and right hand circularly polarized (LCP and RCP)  waves, perpendicular to the wave vector, $k_{l}=\omega_{l}/c$ and 
$V$ is an arbitrary
quantization volume. Then the field-matter interaction involves a coupling term of the form
 \begin{eqnarray}
  \begin{aligned}
  \mathcal{E}_{i}(\omega_{l})&=-\dfrac{i}{\hbar}\sum_{\lambda=1,2}\bra{e_{i}}\mu^{ge}_{i}\ket{g_{i}}.\boldsymbol{\epsilon}^{(\lambda)}_{l}e^{i(\bold{k}_{l}.\bold{R}_{i}-\omega_{l}t)}
    =\mathcal{E}_{il}e^{i(\bold{k}_{l}.\bold{R}_{i}-\omega_{l}t)}
  \label{f_w}
  \end{aligned}
  \end{eqnarray}
  where $\mathcal{E}_{il}=-i\sqrt{\dfrac{\omega_{l}}{2\hbar \epsilon_{0}V}}\times\boldsymbol{\mu}^{ge}_{i}.\boldsymbol{e}^{(\lambda)}$
is a slowly varying function of the virtual photon frequency transferring energy from donor to the acceptors.
For the sake of simplicity,
we assume
a single polarization, so the creation/annihilation operators depend only on the frequency $a(\omega_{l}) = a_{l}$ and the interaction potential is given by
\begin{eqnarray}
  \begin{aligned}
    V_{int}&=\hbar b^{\dagger}_{1}\sum_{l}\mathcal{E}_{1}(\omega_{l}) a_{l}+\hbar b^{\dagger}_{2}\sum_{l}\mathcal{E}_{2}(\omega_{l}) a_{l}+H.c..
  \label{V_int}
  \end{aligned}
  \end{eqnarray}
We proceed by defining the initial state of the donor as $\rho_{0}=\ket{\phi}\bra{\phi}$ ($\rho_{0;kk',qq'}=\bra{1_{k},1_{q}}\rho_{0}\ket{1_{k'},1_{q'}}$) and the initial state of the acceptors as $\ket{g_{1}g_{2}}\bra{g_{1}g_{2}}$. Note that the initial state of the donor can be defined as the pure two-photon state; $
  \ket{\phi}=\sum_{k,q}\eta_{k,q} \ket{1_{k},1_{q}}
  $ or a mixed state in
its spectral decomposition form. The characteristics of the whole system are then determined by  $\ket{g_{1}g_{2}}\bra{g_{1}g_{2}}\otimes \rho_{0}$ and a unitary evolution super operator $U(t)$, since we assume the two acceptors have infinite excited state lifetimes. This informs us about the dynamics of the system over time using the density matrix of the entire system at any time $t$ denoted by $\tilde{\rho}(t)=U(t)\ket{g_{1}g_{2}}\bra{g_{1}g_{2}}\otimes \rho_{0}$.

Generally a pure two-photon state in Hilbert
space is represented by
\begin{eqnarray}
  \begin{aligned}
  \ket{II,donor}&=\sum_{k,q}\eta(\omega_{k},\omega_{q})\ket{1:\omega_{k},\alpha;1:\omega_{q},\beta}
  \label{II_Source}
  \end{aligned}
  \end{eqnarray}
  The symbol $\ket{1_{k},1_{q}}$ represents the tensor product $\ket{1_{k}}  \otimes \ket{1_{q}}$
of two single photon states in frequency mode $\omega_{k(q)}$ of subsystem $\alpha(\beta)$ with the normalized coefficient of $\eta(\omega_{k},\omega_{q})\equiv \eta_{k,q}$.

We then define the donor (emitter) as a three-level particle generating a photon pair through the cascade process (see \autoref{Fig:A1A2Source}).
In this process, the donor is initially excited at $t = 0$ to the top level  $\ket{e}$ with the energy $\hbar(\omega_{\alpha}+\omega_{\beta})$ and width $\gamma_{\alpha}$.
 The first photon is radiated after the transition from $\ket{e}$ to the intermediate state $\ket{m}$ with the frequency $\omega_{\alpha}$ and a Lorentzian distribution in frequency in which its width is $|\gamma_{\alpha}-\gamma_{\beta}|$. For $\gamma_{\alpha}>\gamma_{\beta}$
 there will be some population accumulation, but if $\gamma_{\alpha}\ll \gamma_{\beta}$, the state $\ket{m}$  has a short lifetime and another photon is quickly emitted. At a given time $t$, the state after these emissions is given by\cite{Scully}
\begin{eqnarray}
  \begin{aligned}
  \ket{\Romannum{2},Cas}&=\sum_{k,q}\eta^{Cas}_{k,q} \ket{1_{k},\alpha;1_{q},\beta}\\
  \eta^{Cas}_{k,q}&=\dfrac{\mathcal{N}}{(\omega_{k}-\omega_{\alpha})+i(\gamma_{\alpha}-\gamma_{\beta})}
  \left \{
  \dfrac{1-e^{-\gamma_{\beta}t+i(\omega_{q}-\omega_{\beta})t}}{\omega_{q}-\omega_{\beta}+i\gamma_{\beta}}    - \dfrac{1-e^{-i(\omega_{\alpha}+\omega_{\beta}-\omega_{k}-\omega_{q})t-\gamma_{\alpha}t}}{-(\omega_{\alpha}+\omega_{\beta}-\omega_{k}-\omega_{q})+i\gamma_{\alpha}}
  \right \}
  \label{II_Cas}
  \end{aligned}
  \end{eqnarray}
 In the above expression, $\mathcal{N}$ is the normalization of the two-photon state defined as $\mathcal{N}=\dfrac{2c^{3}\sqrt{\gamma_{\alpha} \gamma_{\beta}}}{V}$ \cite{GRYNBERG_IQO,Scully} associated with the spontaneous emission rate $\gamma_{\alpha(\beta)}=\dfrac{d_{\alpha(\beta)}^{2}\omega_{\alpha(\beta)}^{3}}{6\pi\epsilon_{0}\hbar c^{3}}$\cite{Scully}. Here $d_{\alpha}=\mu^{em,\omega_{\alpha}}$ and $d_{\beta}=\mu^{mg,\omega_{\beta}}$ corresponds to the transition dipole between $\ket{e}\rightarrow\ket{m}$ and $\ket{m}\rightarrow\ket{g}$ respectively.

Note that \autoref{II_Cas} indicates that the state of the entangled photons cannot be factorized into two separable parts. The first term in the
bracket represents the general single photon emission process and the second term corresponds to the frequency anti-correlated (of the second) emission. The (anti-)correlation term comes from energy conservation, since the total energy $\hbar(\omega_{k}+\omega_{q})$ of a photon pair should be close to $\hbar(\omega_{\alpha}+\omega_{\beta})$. 

Classically, there should be four possible ways of passing the two-photon
energy from a single donor to two acceptors: two pathways corresponding to which
photon is absorbed first, and two pairings corresponding
to which acceptor absorbs which photon.
Indeed we  have a temporally
entangled field from the cascade state  with four contributions from
the joint excitation amplitude. However with the entangled input field,
a time ordering is imposed at the emitter and two of the interfering pathways in each acceptor-field pairing have zero amplitude\cite{Inducing-Disallowed_PRL2004}. Later we use the conclusion of this discussion to obtain the correct expression for the probability transition amplitude of the system with entangled photons.

Given the above expression for the field states, the initial and final states of the system are described as 
\begin{eqnarray}
  \begin{aligned}
  \ket{i} &\propto \sum_{k,q}\eta_{k,q} \ket{g_{1}g_{2}}\ket{1_{k},1_{q}},
  \\
  \ket{f} &\propto \ket{e_{1}e_{2}}\ket{0}.
  \label{S_i}
  \end{aligned}
  \end{eqnarray}
This means both acceptors are initially in the ground state, $\ket{g_{1}g_{2}}$,
and the field is initially in a pure two-photon state (or it can be a mixed state) $\sum_{k,q}\eta_{k,q} \ket{1_{k},1_{q}}$. 

From second-order perturbation theory, the state of the acceptor
after two interaction events (at times $t_    {1} < t_{2}$) is obtained from
\begin{eqnarray}
  \begin{aligned}
  U(t) \ket{\phi} \propto \sum_{k,q}\eta_{k,q} \int_{0}^{t}dt_{2} \int_{0}^{t_{2}}dt_{1}V^{I}_{int}(t_{2})V^{I}_{int}(t_{1}) \ket{g_{1}g_{2}}\ket{1_{k},1_{q}}.
  \label{U}
  \end{aligned}
  \end{eqnarray}
The donor-field coupling in the interaction picture, $V^{I}_{int}(t)=e^{iH_{0}t}V_{int}(t)e^{-iH_{0}t}$ can be simplified using the Baker-Campbell-Hausdorff formula.
The joint-excitation amplitude is then given by\cite{Inducing-Disallowed_PRL2004,Scully}
\begin{eqnarray}
  \begin{aligned}
  M(\omega_{1},\omega_{2},t)&=\left \{
  \bra{e_{1}e_{2}}\bra{0}\right \}
  U(t) \ket{\phi},\\
 &=\sum_{k,q} \eta_{k,q}\bra{e_{1}e_{2},0}U(t)\ket{1_{k},1_{q};g_{1}g_{2}}, 
  \label{Amp0}
  \end{aligned}
  \end{eqnarray}
 Assuming the interactions between entangled field and two acceptors are weak, the leading term from the evolution operator \autoref{U} (the second term of Dyson's series) is presented as
 \begin{eqnarray}
  \begin{aligned}
  \bra{e_{1}e_{2}} 
  U^{(2)}(t) \ket{g_{1}g_{2}}&=e^{-iH_{0}t/\hbar}\sum_{k,q}a_{k}a_{q}R_{kq},
   \label{U2}
  \end{aligned}
  \end{eqnarray}
Suppose we have a continuous frequency distribution of the emitted field, so that we can make the replacement
$\sum_{kq} \leftrightarrow 2\int 2\int \Big(\dfrac{V}{(2\pi c)^{3}}\Big)^{2}\omega_{k}^{2}\omega_{q}^{2}d \omega_{k}d \omega_{q}d\Omega_{k}d\Omega_{q}$. Then we arrive at
 \begin{eqnarray}
  \begin{aligned}
  \bra{e_{1}e_{2}} 
  U^{(2)}(t) \ket{g_{1}g_{2}}=\dfrac{V^{2}}{\pi^{4} c^{6}}\int \int d \omega_{k}d \omega_{q} \Big[\omega^{2}_{k}\omega^{2}_{q} a(\omega_{k})a(\omega_{q})R_{kq}\Big]
  \label{U2}
  \end{aligned}
  \end{eqnarray}
 The response function of the uncoupled acceptors to the incoming field $R_{kq}$ is defined as the product of two individual single-photon single-particle response functions \begin{eqnarray}
  \begin{aligned}
  R_{kq}=\mathcal{E}_{1}(\omega_{k})\mathcal{E}_{2}(\omega_{q})e^{-i(\bold{k}_{1}.\bold{R}_{1}+\bold{k}_{2}.\bold{R}_{2})}
  \dfrac{1-e^{i(\omega_{1}-\omega_{k})t}}{\omega_{k}-\omega_{1}}\times
  \dfrac{1-e^{i(\omega_{2}-\omega_{q})t}}{\omega_{q}-\omega_{2}}
  \label{ResFun}
  \end{aligned}
  \end{eqnarray}

If we ignore the propagation length from donor to acceptor by setting $z_{1}\sim z_{2}\sim 0$, it then leads to the approximation $\mathcal{E}_{i}(\omega_{l}) \approx \mathcal{E}_{i}(\omega_{i})$. (Later in this derivation we will include for the propagation length explicitly.) 

Imposing the time ordering $t_{2}>t_{1}$, the resulting expression for the total  joint-excitation amplitude becomes \cite{Inducing-Disallowed_PRL2004}
\begin{eqnarray}
  \begin{aligned}
  \mathcal{M}(\omega_{1},\omega_{2},t)&=M_{\alpha \beta, 12}+M_{\alpha \beta, 21}\\
  &=\dfrac{V^{2}}{\pi^{4} c^{6}}\int \int d \omega_{k}d \omega_{q} \big[\omega^{2}_{k}\omega^{2}_{q} a(\omega_{k})a(\omega_{q})R_{kq}\big](\eta_{kq}+\eta_{qk})
  \label{Mtot1}
  \end{aligned}
  \end{eqnarray}
 
where $M_{\alpha \beta, 12}$ is defined as the probability amplitude that acceptor $1$ interacts with virtual photon $\alpha$ first (at time $t_{1}$), and acceptor $2$ absorbs
virtual photon $\beta$ after that (at time $t_{2}$). This \textit{time-ordered} excitation amplitude is closely related to the two-photon correlation
amplitude. 

Furthermore, the time dependent two-photon excitation probability due to quantum entanglement is defined as the projection (measurement) of the density matrix of the whole system at any time t;
 $\tilde{\rho}=U\ket{g_{1}g_{2}}\bra{g_{1}g_{2}}\otimes \rho_{0}$
onto $\ket{e_{1}e_{2}}\bra{e_{1}e_{2}}$, 
  \begin{eqnarray}
  \begin{aligned}P=Tr \bra{e_{1}e_{2}}\tilde{\rho}\ket{e_{1}e_{2}}=Tr \bra{e_{1}e_{2}}(U\ket{g_{1}g_{2}}\bra{g_{1}g_{2}}\rho_{0})\ket{e_{1}e_{2}}=|\mathcal{M}(\omega_{1},\omega_{2},t)|^{2} 
   \end{aligned}
  \end{eqnarray}
  The "Tr" stands for the trace operation over the field variable.
  Substituting the response function \autoref{ResFun}, into \autoref{Mtot1}, and employing the residue theorem to integrate over the frequencies $\omega_{k}$ and $\omega_{q}$, the total transition
probability amplitude can be determined (more details in SI). However for two-photon cascade state introduced in \autoref{II_Cas}, if $\gamma_{\alpha} \ll \gamma_{\beta}$  the decaying terms related to $e^{-\gamma_{\beta}t}$  vanish at a short time $t\sim \gamma_{\beta}^{-1}$ and the only remaining term at longer time is
\begin{eqnarray}
  \begin{aligned}
  \ket{\Romannum{2},Cas}\approx\dfrac{\mathcal{N}}{(\omega_{k}-\omega_{\alpha})+i(\gamma_{\alpha}-\gamma_{\beta})}
  \left \{
     \dfrac{1-e^{-i(\omega_{\alpha}+\omega_{\beta}-\omega_{k}-\omega_{q})t-\gamma_{\alpha}t}}{(\omega_{\alpha}+\omega_{\beta}-\omega_{k}-\omega_{q})-i\gamma_{\alpha}}
  \right \}
  \label{II_Cas1}
  \end{aligned}
  \end{eqnarray}
Using \autoref{ResFun} and \autoref{Mtot1} and some rearrangement, the transition probability amplitude reads as follows
\begin{eqnarray}
  \begin{aligned}
  \mathcal{M}_{Cas}(\omega_{1},\omega_{2},t)&=-\Lambda
    \Big(
  \dfrac{1}{\omega_{\alpha 1}+i(\gamma_{\beta}-\gamma_{\alpha})}
  +\dfrac{1}{\omega_{\alpha 2}+i(\gamma_{\beta}-\gamma_{\alpha})}
  \Big)
  \left \{
    \dfrac{1-e^{-(\gamma_{\alpha}+i\delta)t}}{\delta-i\gamma_{\alpha}}
  \right \}\\
  &\approx-\Lambda
    \Big(
  \dfrac{1}{\omega_{\alpha 1}+i\gamma_{\beta}}
  +\dfrac{1}{\omega_{\alpha 2}+i\gamma_{\beta}}
  \Big)
  \left \{
    \dfrac{1-e^{-(\gamma_{\alpha}+i\delta)t}}{\delta-i\gamma_{\alpha}}
  \right \}
  \label{Mtot2}
  \end{aligned}
  \end{eqnarray}
where $\Lambda =\mathcal{N}\dfrac{4V^{2}}{\pi^{2} c^{6}}\omega^{2}_{1}\omega^{2}_{2}\mathcal{E}_{1}(\omega_{1})\mathcal{E}_{2}(\omega_{2})
=\dfrac{8V\sqrt{\gamma_{\alpha} \gamma_{\beta}}}{\pi^{2} c^{3}}\omega^{2}_{1}\omega^{2}_{2}\mathcal{E}_{1}(\omega_{1})\mathcal{E}_{2}(\omega_{2})$ and  $\omega_{ij}=\omega_{i}-\omega_{j}$. The detuning for two-entangled virtual photon absorption
where none of the two photons are in resonance with the two acceptors is defined by $\delta=\omega_{\alpha}+\omega_{\beta}-\omega_{1}-\omega_{2} $. Note that in the resonance condition the sum of
their two energies almost matches the sum of the two acceptor's excitation energies; $\omega_{\alpha}+\omega_{\beta}\sim \omega_{1}+\omega_{2} $ (i.e. $\delta \approx 0$).
It then follows from the above expression that the transition probability is given by
\begin{eqnarray}
  \begin{aligned}
  \mathcal{P}_{Cas}(t)&=|\mathcal{M}_{Cas}(\omega_{1},\omega_{2},t)|^{2}
   = | \Lambda|^{2}
    \dfrac{(\omega_{\alpha1}+\omega_{\alpha2})^{2}+4\gamma_{\beta}^{2}}{(\omega_{\alpha1}^{2}+\gamma_{\beta}^{2})(\omega_{\alpha 2}^{2}+\gamma_{\beta}^{2})}
   \Big |  \dfrac{1-e^{-(\gamma_{\alpha}+i\delta)t}}{\delta-i\gamma_{\alpha}}  \Big |^{2}
    \label{Mtot2_delta1}
  \end{aligned}
  \end{eqnarray}
  In the case where $\gamma_{\alpha}\ll \delta$, \autoref{Mtot2_delta1} is recast as
  \begin{eqnarray}
  \begin{aligned}
  \mathcal{P}_{Cas}(t)
  &=  | \Lambda|^{2}
  \dfrac{(\omega_{\alpha1}+\omega_{\alpha2})^{2}+4\gamma_{\beta}^{2}}{(\omega_{\alpha1}^{2}+\gamma_{\beta}^{2})(\omega_{\alpha 2}^{2}+\gamma_{\beta}^{2})}  \times  \dfrac{\sin^{2}(E_{\delta} t/2\hbar)}{(E_{\delta}/2\hbar) ^{2}}
  \quad \quad \gamma_{\alpha}\ll \delta
    \label{Psin}
  \end{aligned}
  \end{eqnarray}
  
    where $E_{\delta}=E_{\omega_{\alpha}+\omega_{\beta}}-E_{\omega_{1}+\omega_{2}}=\hbar \delta$. The above expression shows how the transition probability varies with line-shape parameter $\delta$ and time. At large enough time $\tau$ ($\tau\geq 2 \pi \hbar/E_{\delta} $),  the total probability of transition is a linear function of time and using Fermi's Golden rule accordingly (more details in SI), we can determine the rate as
    
      \begin{eqnarray}
  \begin{aligned}
    W^{Cas}_{\tau}
    &=\lim_{\tau \to\infty} \dfrac{P_{Cas}(\tau)}{\tau}=2\pi\hbar
  \dfrac{(\omega_{\alpha1}+\omega_{\alpha2})^{2}+4\gamma_{\beta}^{2}}{(\omega_{\alpha1}^{2}+\gamma_{\beta}^{2})(\omega_{\alpha 2}^{2}+\gamma_{\beta}^{2})}
   | \Lambda|^{2} \times \delta(E_{\delta})\\
  W^{Cas}_{\tau}&= K_{E2P} |\mathcal{E}_{1}(\omega_{1})\mathcal{E}_{2}(\omega_{2})|^{2}\delta(E_{\delta})
    \label{RateE2P}
  \end{aligned}
  \end{eqnarray}
  
  Note that $\delta(E)$ stands for the Dirac delta function in this formula, and is not to be confused with the line-shape parameter defined above. The E2P coefficient, $K_{E2P}$, is defined as:
  \begin{eqnarray}
  \begin{aligned}
 K_{E2P}&= \dfrac{128 \hbar V^{2}}{\pi^{3} c^{6}}\times
 \omega^{2}_{1}\omega^{2}_{2}
 \times
 \gamma_{\alpha}\gamma_{\beta}
 \times
 \dfrac{(\omega_{\alpha1}+\omega_{\alpha2})^{2}+4\gamma_{\beta}^{2}}{(\omega_{\alpha1}^{2}+\gamma_{\beta}^{2})(\omega_{\alpha 2}^{2}+\gamma_{\beta}^{2})}
 \quad \quad \gamma_{\alpha}\ll \delta
    \label{Rate2in}
  \end{aligned}
  \end{eqnarray}

 The above expression for the energy transfer rate is one of the main accomplishments of this work. This shows that the rate  of the E2P-RET process functions in many respects like the usual one-photon RET rate (see also \autoref{RateE2P}). In particular, we see that the rate depends on the product of decay rates $\gamma_{\beta}\gamma_{\alpha}$, which is related to the rate at which the two photons are emitted.  Also, in the limit where $\omega_{\alpha1}$ and $\omega_{\alpha2}$ are much greater than $\gamma_{\beta}$ we see an inverse dependence on these frequency differences. This inverse dependence means inhibition of the rate as the transitions are detuned from resonance, as makes sense.  Finally note that the volume-dependent terms cancel in the overall rate expression, as is physically required.
 A discussion of what it means to take the time $t$ long enough compared with the lifetime of the two transitions, $t \gg \gamma^{-1}_{\beta}$,$\gamma^{-1}_{\alpha}$ can be found in the SI.
   
RET rate calculations for separated two photons (S2P) may exemplify further details on the features and benefits of using E2P. If we follow the same procedure as above, we are able to derive the same type of expression for the RET rate with two separated photon wave packets of mean frequencies 
$\omega_{\alpha}$ 
and $\omega_{\beta}$, and the corresponding spectral widths $\gamma_{\alpha}$  and $\gamma_{\beta}$. For two identical acceptors ($\omega_{1,2}\approx\omega_{0}$), the resulting expression for S2P-transition probability (See SI for more details) is
 \begin{eqnarray}
  \begin{aligned}
  \mathcal{P}_{11}(\omega_{0},t)=|\mathcal{M}_{11}(\omega_{0},t)|^{2}
  &\approx|\Lambda|^{2} \times
  \Big|
    \dfrac{1-e^{-\gamma_{\alpha}t-i\omega_{\alpha0} t}}{-\omega_{\alpha0}+i\gamma_{\alpha}}\times\dfrac{1-e^{-\gamma_{\beta}t-i\omega_{\beta0} t}}{-\omega_{\beta0}+i\gamma_{\beta}}
  \Big| ^{2}
  \label{P11_delt}
  \end{aligned}
  \end{eqnarray}
  
\begin{figure}
\includegraphics[clip,width=0.49\columnwidth]{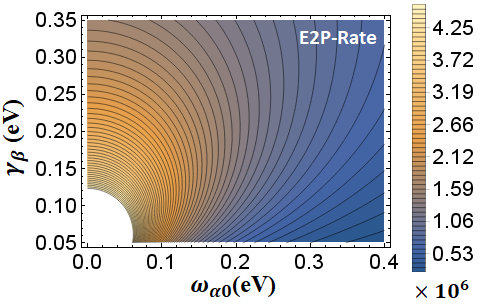}
\includegraphics[clip,width=0.49\columnwidth]{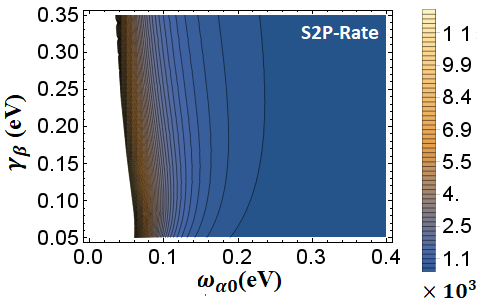}
\caption{Contour-plot of rate for E2P-RET(left) and S2P-RET(right) for identical acceptors in resonance condition. In this set of calculations $\gamma_{\alpha}= 0.005$~eV and $0.05<\gamma_{\beta}<0.35 $~eV, and $\omega_{\alpha 0}=\delta-\omega_{\beta0}$. }
\label{Fig:CompareRate}
\end{figure}

  We numerically calculated the S2P-RET rate in the resonance condition and plotted the result in order to compare with the corresponding E2P-RET rate based on \autoref{Mtot2_delta1} (see \autoref{Fig:CompareRate}).
  We performed our calculations using parameters in the same range as polarization-entangled
photon pairs from the biexciton cascade of a single InAs QD embedded
in a GaAs/AlAs planar microcavity \cite{Cascade_Semicon_NJPhys2006,Cascade_Semicon_Nat2006}. In those experiments the pair entangled photon emissions occur at $1.398$~eV and $1.42$~eV.
   The graphs in \autoref{Fig:CompareRate} show us that  the rate from E2P-RET  is in general  more than three orders of magnitude higher than the rate from S2P-RET. According to the E2P rate graph, the rate has a strong inverse dependence on both $\gamma_{\beta}$ and $\omega_{\alpha0}$ when  $\omega_{\alpha0}$ is close to zero. However by increasing the detuning excitation energy $\omega_{\alpha0}$,  $\gamma_{\beta}$ can effectively control the rate. On the other hand the S2P-RET rate is weakly dependent on $\gamma_{\beta}$ for a wide range of $\omega_{\alpha0}$ values.  These differences arise from the quite different denominators in \autoref{Mtot2_delta1} and the corresponding equation for S2P-RET (see \autoref{P11_delt}).
  
  Note that in the case of S2P-RET, the transition probability is not an explicit  function of the two-photon two acceptor detuning $\delta=\omega_{\alpha}+\omega_{\beta}- \omega_{1}-\omega_{2}$, while it is for E2P-RET.  This plays an important role in determining the  differences between E2P-RET and S2P-RET, and it reflects the crucial influence of the frequency anticorrelation that is built into entanglement.
 
 In \autoref{ResFun} and thereafter, we ignored the propagation length in our derivation. However, in principal the distance dependence of RET in the dipole approximation regime, is contained in matrix elements of the electric dipole-dipole coupling tensor, $\overleftrightarrow {\Theta}(\omega,,\bold{r}_{D},\bold{r}_{A_{i}})$ \cite{Andrews2004_Virtual,ResonanceenergyAndrews2003} that is hidden in the "$\Lambda$
   " coefficient of \autoref{Mtot2}.
 
  Recalling that $\mu^{em,\omega_{\alpha}}(\bold{r}_{D})$, $\mu^{mg,\omega_{\beta}}(\bold{r}_{D})$ and $\mu^{ge}(\textbf{r} _{A_{i}})$ are transition dipoles of the donor and acceptors respectively, we revise the transition amplitude \autoref{Mtot2} to a more comprehensive expression for E2P-RET,
  
 \begin{eqnarray}
  \begin{aligned}
  \mathcal{M}_{Cas}(\omega_{1},\omega_{2},t)&=
  \mathcal{N}\dfrac{4V^{2}}{\pi^{2} c^{6}}\omega^{2}_{1}\omega^{2}_{2}
 \big[ \mu^{em,\omega_{\alpha}}(\bold{r}_{D})
  \overleftrightarrow {\Theta}(\omega_{\alpha},\bold{r}_{D},\bold{r}_{A_{1}})
  \mu^{ge,\omega_{1}}(\bold{r}_{A_{1}})\big]\\
  &\big[ \mu^{mg,\omega_{\beta}}(\bold{r}_{D})
  \overleftrightarrow {\Theta}(\omega_{\beta},\bold{r}_{D},\bold{r}_{A_{2}})
  \mu^{ge,\omega_{2}}(\bold{r}_{A_{2}})\big]
  \times 
  \Big(
  \dfrac{1}{\omega_{\alpha 1}+i\gamma_{\beta}}
  +\dfrac{1}{\omega_{\alpha 2}+i\gamma_{\beta}}
  \Big)
  \left \{
    \dfrac{1-e^{-(\gamma_{\alpha}+i\delta)t}}{\delta-i\gamma_{\alpha}}
  \right \}
  \label{Mtot_r1}
  \end{aligned}
  \end{eqnarray}
 
  Although the distance and orientation dependence of the energy transfer rate are nicely included in the above equation, the time-domain electrodynamics (TED)-RET formulation  \cite{Plasmon-coupled_Wendu2017,Plasmon-coupled_Wendu2017-2} is more convenient for a wide spectrum of applications in inhomogeneous absorbing and dispersive media. Employing this scheme in the dipole approximation, the transition amplitude can be formulated based on the induced field emerging from donor at the position of the two acceptors\cite{Enhanced2PA_2009},
  \begin{eqnarray}
  \begin{aligned}
  \mathcal{M}_{Cas}(\omega_{1},\omega_{2},t)&=
  \mathcal{N}\dfrac{4V^{2}}{\pi^{2} c^{6}}\omega^{2}_{1}\omega^{2}_{2}
 \big[ \mu^{em,\omega_{\alpha}}(\bold{r}_{D})
  \mu^{ge,\omega_{1}}(\bold{r}_{A_{1}})
  \dfrac{\textbf{e}_{A_{1}} .\textbf{E}^{D}(\textbf{r} _{A_{1}},\omega_{\alpha} )}{p_{ex}(\omega_{\alpha})}  \big]\\
  &\big[ \mu^{mg,\omega_{\beta}}(\bold{r}_{D})
    \mu^{ge,\omega_{2}}(\bold{r}_{A_{2}})
    \dfrac{\textbf{e}_{A_{2}} .\textbf{E}^{D}(\textbf{r} _{A_{2}},\omega_{\beta} )}{p_{ex}(\omega_{\beta})}\big]\\
  &\times
  \Big(
  \dfrac{1}{\omega_{\alpha 1}+i\gamma_{\beta}}
  +\dfrac{1}{\omega_{\alpha 2}+i\gamma_{\beta}}
  \Big)
  \left \{
    \dfrac{1-e^{-(\gamma_{\alpha}+i\delta)t}}{\delta-i\gamma_{\alpha}}
  \right \}
  \label{Mtot_r}
  \end{aligned}
  \end{eqnarray}
  
Here, $\textbf{E}^{D}(\textbf{r} _{A_{i}},\omega_{\alpha ( \beta)})$ (i=1,2) is the induced electric field mode with angular
frequency $\omega_{\alpha ( \beta)}$ and unit polarization vector $\textbf{e}_{A_{i}}$ and at the position of acceptor $\bold{r}_{A_{i}}$. We proceed by using the above concept to obtain a general expression for the  E2P-RET rate in a manner similar to \autoref{RateE2P};
 \begin{eqnarray}
  \begin{aligned}
W^{Cas}_{\tau}&= K_{E2P} |\Pi|^{2}
\Big | \dfrac{\textbf{e}_{A_{1}} .\textbf{E}^{D}(\textbf{r} _{A_{1}},\omega_{\alpha} )}{p_{ex}(\omega_{\alpha})}\quad  \dfrac{\textbf{e}_{A_{2}}.\textbf{E}^{D}(\textbf{r} _{A_{2}},\omega_{\beta} )}{p_{ex}(\omega_{\beta})}\Big|^{2}
   \delta(E_{\delta})
    \label{RateE2P_R}
  \end{aligned}
  \end{eqnarray}
  where
  $
 \Pi=
  \mu^{em,\omega_{\alpha}}
  \mu^{ge,\omega_{1}} \mu^{mg,\omega_{\beta}}
    \mu^{ge,\omega_{2}}
  $. 
  The above expression is another major accomplishment in this work which shows a fourth power dependence (compared with single photon energy transfer) of the rate on the induced electric fields by the donor. This expression allows us to calculate the enhanced E2P-RET rate in arbitrary media, which is more convenient for practical implementation.  We also note that the dependence of the energy transfer rate on polarization properties of the donor and acceptors in included in this formula.
  
 In a regime where the size of the donor and acceptor are as big as the distance between them, we should go beyond the point dipole approximation by including the effect of higher order multipoles. In that case, the E2P-RET rate is calculated according to the total induced field in the system (more details in SI) using the following expression, 
    \begin{eqnarray}
  \begin{aligned}
W^{Cas}_{\tau}&= K_{E2P}
\Big |
    \mathcal{E}^{Total}(\textbf{r}_{A_{1}},\textbf{r}_{D},\omega_{\alpha})\Big| ^{2}
   \quad \Big | \mathcal{E}^{Total}(\textbf{r}_{A_{2}},\textbf{r}_{D},\omega_{\beta})
   \Big| ^{2}\delta(E_{\delta})
    \label{RateE2P_R}
  \end{aligned}
  \end{eqnarray}
 The proposed theory provides a framework for simulation that has
significant computational advantages compared to calculating
the coupling factor utilizing dyadic Green’s functions.

In conclusion, we developed a new theory for the resonance energy transfer between a donor and a pair of uncoupled acceptor  particles via entangled photons. The underlying mechanism is uncovered through the joint  excitation of acceptors using a temporally  entangled  field. The calculated result shows more than a three order of magnitude enhancement in the E2P-RET rate compared with the S2P case for parameters that are relevant to biexciton sources.
Empowered with the quantum description of light, our theory provides a way to control the E2P-RET phenomena through the effective coefficient $K_{E2P}$. This coefficient emphasizes the importance of the emission rate parameters of the donor and there was also an important effect arising from detuning of the emitted photons energy relative to the excitation energies of the acceptors.
Furthermore we have extended our theory to include for the effect of donor-acceptor separation, and the influence of inhomogeneous, dispersive and absorptive materials with any space-dependent, frequency-dependent dielectric function and with any size of donor and acceptor.

 Since SPDC is a very common method for producing entangled photons in  experimental studies, it will also be important to examine energy transfer associated with SPDC sources and comparing it with the cascade source of the present study. 
The theoretical results of this work will lead the way to a new platform for exploring exciton and biexciton transport in coupled plasmonic-semiconductor nanostructures, with potential applications in spectroscopy, nanophotonics devices, biosensing  and quantum information.
\section{Supporting Information}
Details of theoretical derivations can be found here. 

\begin{acknowledgement}
This work was supported by the U.S. National Science Foundation under Grant No. CHE-1760537. This research was supported in part through the computational resources and staff contributions provided for the Quest high performance computing facility at Northwestern University which is jointly supported by the Office of the Provost, the Office for Research, and Northwestern University Information Technology. 
\end{acknowledgement}

\bibliography{Cites}
\newpage

\section*{Supporting Information: Entangled Photon Resonance Energy Transfer in Arbitrary Media}



\renewcommand*{\equationautorefname}{Eq.}
\renewcommand*{\figureautorefname}{Fig.}
\renewcommand{\thefigure}{S\arabic{figure}}
\renewcommand{\theequation}{S\arabic{equation}}
\renewcommand{\thepage}{S\arabic{page}}
\setcounter{figure}{0}
\setcounter{equation}{0} 

  \section*{General RET Rate Calculation}
  The total transition
probability amplitude using Eq.15 in main text reads  as follows
\begin{eqnarray}
  \begin{aligned}
  \mathcal{M}_{Cas}(\omega_{1},\omega_{2},t)=
  \mathcal{N}\dfrac{4V^{2}}{\pi^{2} c^{6}}\omega^{2}_{1}\omega^{2}_{2}
  \dfrac{\mathcal{E}_{1}(\omega_{1})\mathcal{E}_{2}(\omega_{2})}{(\omega_{\beta 2}-\delta)-i(\gamma_{\beta}-\gamma_{\alpha})}
  \left \{
  \dfrac{1-e^{-(\gamma_{\beta}+i\omega_{\beta 2})t}}{\omega_{\beta 2}-i\gamma_{\beta}}-
  \dfrac{1-e^{-(\gamma_{\alpha}+i\delta)t}}{\delta-i\gamma_{\alpha}}
  \right \}+ (1 \leftrightarrow 2)
  \label{Mtot2s}
  \end{aligned}
  \end{eqnarray}
For two identical acceptors ($\omega_{1}=\omega_{2}=\omega_{0}$), the total transition probability is therefore
\begin{eqnarray}
  \begin{aligned}
  \mathcal{P}_{Cas}(t)=
  \dfrac{4|\Lambda|^{2}}{(\omega_{\beta 0}-\delta)^{2}+(\gamma_{\beta}-\gamma_{\alpha})^{2}}
  \Bigg|
  \dfrac{1-e^{-(\gamma_{\beta}+i\omega_{\beta 0})t}}{\omega_{\beta 0}-i\gamma_{\beta}}-
  \dfrac{1-e^{-(\gamma_{\alpha}+i\delta)t}}{\delta-i\gamma_{\alpha}}
  \Bigg|^{2}
  \label{Mtot2_deltas}
  \end{aligned}
  \end{eqnarray}
where $\Lambda =\mathcal{N}\dfrac{4V^{2}}{\pi^{2} c^{6}}\omega^{2}_{1}\omega^{2}_{2}\mathcal{E}_{1}(\omega_{1})\mathcal{E}_{2}(\omega_{2})$ . If the resonance condition ($\delta \sim 0$) is met, the transition probability simplifies to
\begin{eqnarray}
  \begin{aligned}
    \mathcal{P}^{Res}_{Cas}(t)\approx
  |\Lambda|^{2}
  \left \{
    \dfrac{e^{-\gamma_{\beta}t}sin^{2}(\omega_{\beta0}t/2)}{(\omega_{\beta0}^{2}+\gamma_{\beta}^{2})^{2}/4}+
  \dfrac{1}{(\omega_{\beta0}^{2}+\gamma_{\beta}^{2})}
  \left|
  \dfrac{1-e^{-\gamma_{\beta}t}}{\omega_{\beta0}-i\gamma_{\beta}}+ 
   \dfrac{1-e^{-\gamma_{\alpha}t}}{i\gamma_{\alpha}}
  \right|^{2}
    \right \}
  \label{Ptots}
  \end{aligned}
  \end{eqnarray}
  
Although the above expression gives us some insight on the characteristic properties of time-dependent transition probability and associated parameters, it is a very complicated expression if we wish to calculate the rate of  energy transfer initiated by entangled photons. Using concepts that arise in deriving Fermi's Golden rule, a general formula for determining the RET rate can be developed in the limit of large $\tau$ and $\delta \rightarrow 0$,
\begin{eqnarray}
  \begin{aligned}
  W_{\tau}&=\lim_{\tau \to\infty} \dfrac{P_{Cas}(\tau)}{\tau}
  \label{Ratedefs}
   \end{aligned}
\end{eqnarray}
where $E_{\delta}=E_{\omega_{\alpha}+\omega_{\beta}}-E_{\omega_{1}+\omega_{2}}=\hbar \delta$.
In the main text with some assumptions, we give an analytical expression for the RET rate that results from this analysis.

For the
two-photon cascade state introduced in the main text, if $\gamma_{\alpha} \ll \gamma_{\beta}$  the decaying terms related to $e^{-\gamma_{\beta}t}$  vanish at short time $t\sim \gamma_{\beta}^{-1}$,
and if the spectral width $\gamma_{\alpha}$ is much smaller than the two-photon two-acceptor's detuning $\delta$, the transition probability is recast as
  \begin{eqnarray}
  \begin{aligned}
  \mathcal{P}_{Cas}(t)&=
  | \Lambda|^{2}
  \dfrac{(\omega_{\alpha1}+\omega_{\alpha2})^{2}+4\gamma_{\beta}^{2}}{(\omega_{\alpha1}^{2}+\gamma_{\beta}^{2})(\omega_{\alpha 2}^{2}+\gamma_{\beta}^{2})}
  \times
  \dfrac{\sin^{2}(\delta t/2)}{(\delta/2) ^{2}}\\
  &=  | \Lambda|^{2}
  \dfrac{(\omega_{\alpha1}+\omega_{\alpha2})^{2}+4\gamma_{\beta}^{2}}{(\omega_{\alpha1}^{2}+\gamma_{\beta}^{2})(\omega_{\alpha 2}^{2}+\gamma_{\beta}^{2})}  \times  \dfrac{\sin^{2}(E_{\delta} t/2\hbar)}{(E_{\delta}/2\hbar) ^{2}}
  \quad \quad \gamma_{\alpha}\ll \delta
    \label{Psins}
  \end{aligned}
  \end{eqnarray}
  Rearranging \autoref{Psins}, we see for large enough $t=\tau$ and in the resonance condition $\delta \sim 0$, $F(0) \rightarrow \infty$ which means $F$ behaves a like a delta function.
   \begin{eqnarray}
  \begin{aligned}
  F(E_{\delta})= \dfrac{\sin^{2}(E_{\delta} \tau/2\hbar)}{\tau(E_{\delta}/2\hbar) ^{2}}=| \Lambda|^{-2}
  \dfrac{(\omega_{\alpha1}^{2}+\gamma_{\beta}^{2})(\omega_{\alpha 2}^{2}+\gamma_{\beta}^{2})}{(\omega_{\alpha1}+\omega_{\alpha2})^{2}+4\gamma_{\beta}^{2}}
  \times \dfrac{P_{Cas}(\tau)}{\tau}
  \label{Deltas}
  \end{aligned}
  \end{eqnarray}
  has the property  that
 \begin{eqnarray}
  \begin{aligned}
  \int_{-\infty}^{+\infty}F(E_{\delta}) dE_{\delta}=2\pi\hbar
  \label{Deltas}
  \end{aligned}
  \end{eqnarray}
  From this it follows that
  \begin{eqnarray}
  \begin{aligned}
  \lim_{\tau \to\infty} F(E_{\delta})=2\pi\hbar\delta(E_{\delta})
  \label{Ratedefs}
   \end{aligned}
\end{eqnarray}
Thus in this limit, the total probability of transition is a linear function of time and using Fermi's Golden rule accordingly, we can determine the rate as
  \begin{eqnarray}
  \begin{aligned}
  W_{\tau}&=\lim_{\tau \to\infty} \dfrac{P_{Cas}(\tau)}{\tau}=2\pi\hbar
  \dfrac{(\omega_{\alpha1}+\omega_{\alpha2})^{2}+4\gamma_{\beta}^{2}}{(\omega_{\alpha1}^{2}+\gamma_{\beta}^{2})(\omega_{\alpha 2}^{2}+\gamma_{\beta}^{2})}
   | \Lambda|^{2} \times \delta(E_{\delta})
    \label{RateE2Ps}
  \end{aligned}
  \end{eqnarray}

\section*{Very Long Time Transition Probability and Rate}
For time $t$ long compared to the lifetime of the two transitions, $t \gg \gamma^{-1}_{\beta}$,$\gamma^{-1}_{\alpha}$, the two-photon wave-packet given by Eq.7 in the main text is time-independent \cite{Scully, KhanPRA2006, RevModPhys2009}. However, there still exist significant anti-correlations between the emitted fields. As we mentioned earlier, in order to produce a two-entangled photon wavepacket, the excited state's life time should be much longer than the spontaneous
emission rates of the intermediate state $\ket{m}$; i.e. $\gamma_{\beta}\gg \gamma_{\alpha}$.
This leads to the second emission occurring soon after the first one, leading to following state of light, 
\begin{eqnarray}
  \begin{aligned}
  \ket{\Romannum{2},Cas}_{t\rightarrow\infty}&=\sum_{k,q}
  \dfrac{\mathcal{N}}
    {[-(\omega_{\alpha}+\omega_{\beta}-\omega_{k}-\omega_{q})+i\gamma_{\alpha}](\omega_{q}-\omega_{\beta}+i\gamma_{\beta})} \ket{1_{k},1_{q}}
    \label{II_Cas-infinitys}
  \end{aligned}
  \end{eqnarray}
  Using the above expression for the quantum state of light and in the same manner as discussed in the main text, we  can obtain the time-independent transition amplitude and the total transition probability as
  \begin{eqnarray}
  \begin{aligned}
  \mathcal{M}_{t\rightarrow\infty}(\omega_{1},\omega_{2})&=-
  \mathcal{N}\dfrac{4V^{2}}{\pi^{2} c^{6}}\omega^{2}_{1}\omega^{2}_{2}
  \dfrac{\mathcal{E}_{1}(\omega_{1})\mathcal{E}_{2}(\omega_{2})}{\delta-i\gamma_{\alpha}}
  \left \{
  \dfrac{1}{\omega_{\beta 2}-i\gamma_{\beta}}+
  \dfrac{1}{\omega_{\beta 1}-i\gamma_{\beta}}
  \right \}
  \\
  \mathcal{P}_{t\rightarrow\infty}&=|M(\omega_{1},\omega_{2})|^{2}=
  \dfrac{|\Lambda|^{2}}
  {(\delta^{2}+\gamma_{\alpha}^2)}
  \times
  \dfrac{(\omega_{\beta 1}+\omega_{\beta 2})^{2}+4\gamma_{\beta}^{2}}{(\omega_{\beta 1}^{2}+\gamma_{\beta}^{2})(\omega_{\beta 2}^{2}+\gamma_{\beta}^{2})}
  \label{M-infinitys}
  \end{aligned}
  \end{eqnarray}
  When the resonance condition is satisfied ($\delta\sim 0$), the transition probability is then given by
  \begin{eqnarray}
  \begin{aligned}
  \mathcal{P}^{Res}_{Cas}&\approx
      &\approx
  \dfrac{1}{2\pi \hbar\gamma_{\alpha}}
  K_{E2P}\times  |\mathcal{E}_{1}(\omega_{1})\mathcal{E}_{2}(\omega_{2})|^{2}\quad \quad  \gamma_{\alpha}t,\gamma_{\beta}t \gg 1
   \label{P-infinitys}
  \end{aligned}
  \end{eqnarray}
It should be noted that the Fermi's golden rule is valid when the initial state has not been significantly depleted by transferring into the final states. This is related to the population accumulation of the final state, and in the large time limit it leads to a finite probability for populating the excited states of the acceptors. It therefore makes sense that the probability in \autoref{P-infinitys} involves the same dependence on donor/acceptor frequency differences as the rate coefficient in \autoref{Rate2in} (in main text).
  
  \section*{S2P-transition Probability}
  Calculating the RET rate for two separated photons can give us more insight into the features and benefits of using E2P. If we follow the same procedure as above, we are able to derive the same type of expression for the RET rate with two separated photon wave packets of mean frequencies 
$\omega_{\alpha}$ 
and $\omega_{\beta}$, and the corresponding spectral widths $\gamma_{\alpha}$  and $\gamma_{\beta}$.
In this case neither of the two photons are in resonance with the two acceptors but again the sum of their two energies almost matches the sum of the emitted energy from donors; $\omega_{\alpha}+\omega_{\beta}\cong \omega_{1}+\omega_{2} $.
Here we compare the rate of E2P-RET with two virtual photons passing energy to two non-interacting acceptors. We assume that the two separate events happen almost at the same time but otherwise there is no correlation between the two-photons. This should not be mistaken with two-photon energy transfer by a pair of donors and acceptors, i.e. this is not comparable with the two-photon absorption/emission process\cite{AQED_theory_JPC2001,ThreeCenter_Andrews98,Multichromophore_Andrew_JCP2003,2Pfluorescence_AndrewsJCP98}. However each photon has a chance to interact with either acceptor 1 or 2 in our derivation. We assume the donor particle that has two different excited states, emitting two unentangled photons followed by absorption via two acceptors. The near-simultaneous arrival of the two unentangled photons is not crucial in this derivation since we also assume that the acceptors have infinite lifetime.  (Note: if we include for decay of the acceptors into the account then these two events need to occur very close in time as otherwise by the time a photon is received by the second acceptor, the first acceptor can be depleted.) 
 
  To construct nonentangled photons with the same mean energy and the same single photon spectrum, we define the state of the system as two quasimonochromatic uncorrelated photons emitted by two uncorrelated particles excited at the same time earlier and arriving at the acceptor's location at $t =0$ as \cite{GRYNBERG_IQO}
  \begin{eqnarray}
  \begin{aligned}
  \ket{\Psi^{11}}&=\sum_{k} 
g_{\alpha}(\omega_{k})
  \left \{
  \dfrac{1-e^{-\gamma_{\alpha}t-i(\omega_{\alpha}-\omega_{k})t}}{\omega_{k}-\omega_{\alpha}+i\gamma_{\alpha}}  
    \right \}\ket{1_{k}}
\otimes \sum_{q}g_{\beta} (\omega_{q})
  \left \{
   \dfrac{1-e^{-\gamma_{\beta}t-i(\omega_{\beta}-\omega_{q})t}}{\omega_{q}-\omega_{\beta}+i\gamma_{\beta}} 
  \right \}\ket{1_{q}}
  \label{II_psis}
  \end{aligned}
  \end{eqnarray}
where $g_{\alpha(\beta)}(\omega_{k(q)})=\mu_{\alpha(\beta)}\sqrt{\dfrac{ \omega_{k(q)}}{2\hbar\epsilon_{0}V}}$ and $\mu_{\alpha(\beta)}$ is the transition dipole moment of each particle.  We also have  the option of choosing
one of two special cases that will allow for a quantitative evaluation of the role of correlations: the donor emits  correlated but separable photons with the density matrix: $\rho_{1}=\sum_{k,q} \rho_{0;kk,qq}\ket{1_{k},1_{q}}\ket{1_{k},1_{q}}$ or we have a fully factorized state with
$\rho_{2}=\Big(\sum_{k,q'} \rho_{0;k,q'}\ket{1_{k}}\ket{1_{k}}\Big)\otimes \Big(\sum_{k',q} \rho_{0;k',q}\ket{1_{q}}\ket{1_{q}}\Big)$ \cite{S2P_PRL2000}.

Based on \autoref{II_psis} the probability amplitude can be determined through
\begin{eqnarray}
  \begin{aligned}
  \mathcal{M}_{11}(\omega_{1},\omega_{2},t)=
  \Lambda
  \left \{
  \dfrac{1-e^{-\gamma_{\alpha}t-i\omega_{\alpha1} t}}{-\omega_{\alpha1}+i\gamma_{\alpha}} \times
  \dfrac{1-e^{-\gamma_{\beta}t-i\omega_{\beta2} t}}{-\omega_{\beta2}+i\gamma_{\beta}}+1\leftrightarrow2
  \right \}
  \label{Mtot_2ps}
  \end{aligned}
  \end{eqnarray}
  
 And for two identical acceptors ($\omega_{1,2}\approx\omega_{0}$), we can simplify the transition probability equation as
 \begin{eqnarray}
  \begin{aligned}
  \mathcal{P}_{11}(\omega_{0},t)=|\mathcal{M}_{11}(\omega_{0},t)|^{2}
  &\approx|\Lambda|^{2} \times
  \Big|
    \dfrac{1-e^{-\gamma_{\alpha}t-i\omega_{\alpha0} t}}{-\omega_{\alpha0}+i\gamma_{\alpha}}\times\dfrac{1-e^{-\gamma_{\beta}t-i\omega_{\beta0} t}}{-\omega_{\beta0}+i\gamma_{\beta}}
  \Big| ^{2}
  \label{P11_delt}
  \end{aligned}
  \end{eqnarray}
  \section*{Distance-dependence of E2P-RET Process}
  
 In the first part of the derivation in the main text, we ignored the propagation length. However, in principal the distance dependence in RET in the dipole approximation regime comes from
   matrix elements of the electric dipole-dipole coupling tensor, $\overleftrightarrow {\Theta}(\omega,,\bold{r}_{D},\bold{r}_{A_{i}})$ \cite{Andrews2004_Virtual,ResonanceenergyAndrews2003} hidden in the "$\Lambda$
   " coefficient in \autoref{Mtot2} in main text.
   
    Generally, $\overleftrightarrow {\Theta}(\omega,\bold{r}_{D},\bold{r}_{A})$ in vacuum is defined as
 \begin{eqnarray}
  \begin{aligned}
    \overleftrightarrow {\Theta}(\omega,\bold{r}_{D},\bold{r}_{A})=
    \dfrac{\omega^{3}e^{i\omega R/c}}{4 \pi \epsilon_{0}c^{3}}
     \left \{
  (\delta_{mn}-3e_{R_{m}}e_{R_{n}})
  \Big( \dfrac{c^{3}}{\omega^{3}R^{3}}-\dfrac{ic^{2}}{\omega^{2}R^{2}}\Big)
  -(\delta_{mn}-e_{R_{m}}e_{R_{n}})\dfrac{c}{\omega R}
  \right \}
  \label{V_rs}
  \end{aligned}
  \end{eqnarray}
  where $R = |\bold{R}| = |\bold{r}_{A}-\bold{r}_{D}|$ is the amplitude of the spatial displacement
vector between the donor and the acceptor, $e_{R_{i}}$ stands for
the $i$th component of the unit vector of $\bold{R}$ ($\bold{e}_{R} =\bold{R}/R$) and $\delta$ denotes the Kronecker delta.

  Setting $\mu^{em,\omega_{\alpha}}(\bold{r}_{D})$, $\mu^{mg,\omega_{\beta}}(\bold{r}_{D})$ and $\mu^{ge}(\textbf{r} _{A_{i}})$ as the transition dipoles of the donor during the cascade process and also of the acceptors respectively, we revise the transition amplitude to a more comprehensive expression when the process involves E2P,
 \begin{eqnarray}
  \begin{aligned}
  \mathcal{M}_{Cas}(\omega_{1},\omega_{2},t)&=
  \mathcal{N}\dfrac{4V^{2}}{\pi^{2} c^{6}}\omega^{2}_{1}\omega^{2}_{2}
 \big[ \mu^{em,\omega_{\alpha}}(\bold{r}_{D})
  \overleftrightarrow {\Theta}(\omega_{\alpha},\bold{r}_{D},\bold{r}_{A_{1}})
  \mu^{ge,\omega_{1}}(\bold{r}_{A_{1}})\big]\\
  &\big[ \mu^{mg,\omega_{\beta}}(\bold{r}_{D})
  \overleftrightarrow {\Theta}(\omega_{\beta},\bold{r}_{D},\bold{r}_{A_{2}})
  \mu^{ge,\omega_{2}}(\bold{r}_{A_{2}})\big]\\
  &\times \left \{
  \dfrac{1 }
  { \omega_{\beta 2}-\delta-i(\gamma_{\beta}-\gamma_{\alpha})}
  \Big(
  \dfrac{1-e^{-(\gamma_{\beta}+i\omega_{\beta 2})t}}{\omega_{\beta 2}-i\gamma_{\beta}}-
  \dfrac{1-e^{-(\gamma_{\alpha}+i\delta)t}}{\delta-i\gamma_{\alpha}}
  \Big )
  + (1 \leftrightarrow 2)
  \right \}
  \label{Mtot_r1s}
  \end{aligned}
  \end{eqnarray}
  The superscripts $e,m$ and $g$ represent the excited, intermediate, and ground states respectively. 
  
  Although the distance and orientation dependence are nicely included in the above equation, it is more convenient and more general to use the time-domain electrodynamics resonance energy transfer (TED-RET) formulation  \cite{Plasmon-coupled_Wendu2017,Plasmon-coupled_Wendu2017-2} which enables a wide spectrum of applications, particularly in inhomogeneous absorbing and dispersive media using a real-time electrodynamics approach. In this scheme the donor is assumed to be a single radiating particle positioned at $\textbf{r}_{D}$ whose size is much smaller than the distance between the donor and the acceptor. Thus we employ the point-dipole approximation; $\textbf{P}_{ex}(\textbf{r}, t) = \textbf{p}_{ex}(t)\delta (\textbf{r} - \textbf{r}_{D})$, in which the external polarization $\textbf{P}_{ex}$ generated by the donor in a dielectric medium is defined by $\textbf{p}_{ex}(t)$ (or its temporal Fourier transform  $\textbf{p}_{ex}(\omega)$).
Furthermore, the transition matrix element at angular frequency $\omega$, is calculated from the electric field $\textbf{E}^{D}(\textbf{r}_{A})$ at the position of the acceptor ($\textbf{r}_{A}$) originating from the donor as follows
 \begin{eqnarray}
  \begin{aligned}
  \mathcal{M}(\textbf{r} _{D} , \textbf{r} _{A}, \omega)=
  \boldsymbol{\mu}^{eg}(\textbf{r} _{D})\overleftrightarrow {\Theta}(\textbf{r} _{D},\textbf{r} _{A},\omega) \boldsymbol{\mu}^{ge}(\textbf{r} _{A})
  =-\mu^{eg}(\textbf{r} _{A}) \mu^{ge}(\textbf{r} _{D})
  \dfrac{\textbf{e}_{A} .\textbf{E}^{D}(\textbf{r} _{A},\omega )}{p_{ex}(\omega)} 
  \label{TransMats}
  \end{aligned}
  \end{eqnarray}
Employing the above equation, the magnitude of transition dipole of the donor (acceptors) and $\textbf{p}_{ex}(\omega)$ can be obtained via computational electromagnetic software based on the FDTD method. The normalization factor $p_{ex}(\omega)$ is the amplitude in the frequency domain of the Hertzian dipole $\textbf{p}_{ex}$. Dividing by this factor and then multiplying by $\mu^{ge}(\textbf{r}_{D})$, one obtains the correct field strength generated by the donor.
 
  Plugging the right hand side of \autoref{TransMats} in \autoref{Mtot_r1s}, the transition amplitude can be calculated based on the induced field from donor at the position of the two acceptors\cite{Enhanced2PA_2009},
  \begin{eqnarray}
  \begin{aligned}
  \mathcal{M}_{Cas}(\omega_{1},\omega_{2},t)&=
  \mathcal{N}\dfrac{4V^{2}}{\pi^{2} c^{6}}\omega^{2}_{1}\omega^{2}_{2}
 \big[ \mu^{em,\omega_{\alpha}}(\bold{r}_{D})
  \mu^{ge,\omega_{1}}(\bold{r}_{A_{1}})
  \dfrac{\textbf{e}_{A_{1}} .\textbf{E}^{D}(\textbf{r} _{A_{1}},\omega_{\alpha} )}{p_{ex}(\omega_{\alpha})}  \big]\\
  &\big[ \mu^{mg,\omega_{\beta}}(\bold{r}_{D})
    \mu^{ge,\omega_{2}}(\bold{r}_{A_{2}})
    \dfrac{\textbf{e}_{A_{2}} .\textbf{E}^{D}(\textbf{r} _{A_{2}},\omega_{\beta} )}{p_{ex}(\omega_{\beta})}\big]\\
  &\times \left \{
  \dfrac{1 }
  { \omega_{\beta 2}-\delta-i(\gamma_{\beta}-\gamma_{\alpha})}
  \Big (
  \dfrac{1-e^{-(\gamma_{\beta}+i\omega_{\beta 2})t}}{\omega_{\beta 2}-i\gamma_{\beta}}-
  \dfrac{1-e^{-(\gamma_{\alpha}+i\delta)t}}{\delta-i\gamma_{\alpha}}
  \Big )
  + (1 \leftrightarrow 2)
  \right \}
  \label{Mtot_rs}
  \end{aligned}
  \end{eqnarray}
  
Here, $\textbf{E}^{D}(\textbf{r} _{A_{i}},\omega_{\alpha / \beta})$ (i=1,2) is the induced electric field mode with angular
frequency $\omega_{\alpha / \beta}$ and unit polarization vector $\textbf{e}_{A_{i}}$ and at the position of acceptor $\bold{r}_{A_{i}}$. The dipole moments $\mu^{ge,\omega_{i}}(\bold{r}_{A_{i}})$ 
represent an average over random orientations\cite{Enhanced2PA_2009}.
If we just follow the steps that we used to obtain \autoref{RateE2Ps}, the general expression for the  rate of resonance energy transfer is given by
 \begin{eqnarray}
  \begin{aligned}
W^{Cas}_{\tau}&= K_{E2P} |\Pi|^{2}
\Big | \dfrac{\textbf{e}_{A_{1}} .\textbf{E}^{D}(\textbf{r} _{A_{1}},\omega_{\alpha} )}{p_{ex}(\omega_{\alpha})}\quad  \dfrac{\textbf{e}_{A_{2}}.\textbf{E}^{D}(\textbf{r} _{A_{2}},\omega_{\beta} )}{p_{ex}(\omega_{\beta})}\Big|^{2}
   \delta(E_{\delta})
    \label{RateE2P_Rs}
  \end{aligned}
  \end{eqnarray}
  where
  $
 \Pi=
  \mu^{em,\omega_{\alpha}}
  \mu^{ge,\omega_{1}} \mu^{mg,\omega_{\beta}}
    \mu^{ge,\omega_{2}}
  $. 
  
 When the optical transition in donor and/or acceptor is dipole forbidden, or the size of the donor and/or the acceptor is comparable with distance between them, the dipole approximation is not valid anymore and the effect of higher order multipoles needs to be included. 
  In that case, the total electric field in the system is the sum of the E-field of the electric dipole ($ \textbf{E}^{ED} $),
  the E-field of the magnetic dipole ($ \textbf{E}^{MD} $), and the E-field of the electric quadrupole ($ \textbf{E}^{EQ}  $), while the total magnetic field is the sum of the M-field of the electric dipole ($ \textbf{B}^{ED}  $), the M-field of the magnetic dipole ($ \textbf{B}^{MD}  $), and the M-field of the electric quadrupole ($ \textbf{B}^{EQ}  $). Including for these effects,  the transition matrix elements for RET in terms of interactions between acceptor transition multipoles (electric dipole $\mu$, magnetic dipole $m$, and electric quadrupole $\overleftrightarrow{Q}$) and the corresponding electromagnetic fields generated by the donor transition multipoles are given by \cite{BeyondPDApprox_Jpcc2018}: 
  \begin{subequations}
  \label{Eq:Mfinal}
 \begin{align}
 \label{Eq:Mtotal}
 \mathcal{E}^{Total}(\textbf{r}_{A},\textbf{r}_{D},\omega)&=
  \mathcal{E}^{e}(\textbf{r}_{A},\textbf{r}_{D},\omega)+
  \mathcal{E}^{m}(\textbf{r}_{A},\textbf{r}_{D},\omega)+
  \mathcal{E}^{q}(\textbf{r}_{A},\textbf{r}_{D},\omega),
\\
  \label{Eq:Mes}
  \begin{split}
\mathcal{E}^{e}(\textbf{r}_{A},\textbf{r}_{D},\omega)
  &= -\mu^{eg}_{A} \textbf{e}^{\mu}_{A} \cdot
  \left[
    \dfrac{\mu^{ge}_{D}}{p(\omega)} \textbf{E}^{ED}
  + \dfrac{ m^{ge}_{D}}{m(\omega)} \textbf{E}^{MD}
  + \dfrac{ q^{ge}_{D}}{q(\omega)} \textbf{E}^{EQ}
  \right],
  \end{split}
\\
  \label{Eq:Mms}
  \begin{split}
\mathcal{E}^{m}(\textbf{r}_{A},\textbf{r}_{D},\omega)
  &= -m^{eg}_{A} \textbf{e}^{m}_{A} \cdot 
  \left[ 
    \dfrac{\mu^{ge}_{D}}{p(\omega)} \textbf{B}^{ED}
  + \dfrac{m^{ge}_{D}}{m(\omega)} \textbf{B}^{MD}
  + \dfrac{q^{ge}_{D}}{q(\omega)} \textbf{B}^{EQ}
  \right],
  \end{split}
\\
  \label{Eq:Mqs}
  \begin{split}
\mathcal{E}^{q}(\textbf{r}_{A},\textbf{r}_{D},\omega)
  &= -q^{eg}_{A} \overleftrightarrow{e}^{q}_{A} :
  \left[
    \dfrac{\mu^{ge}_{D}}{p(\omega)} \nabla\textbf{E}^{ED}
  + \dfrac{m^{ge}_{D}}{m(\omega)}  \nabla\textbf{E}^{MD}
  + \dfrac{q^{ge}_{D}}{q(\omega)} \nabla\textbf{E}^{EQ}
  \right],
  \end{split}
 \end{align}
\end{subequations}
  where $ p(\omega)$, $ m(\omega)$ and $ q(\omega)$ are the amplitude of the physical multipoles used to calculate the fields generated by the
electric dipole ($\textbf{p}$), magnetic dipole ($\textbf{m}$) and electric quadrupole ($\overleftrightarrow{Q}$), respectively. These normalization factors ensure that the fields have the correct magnitudes corresponding to the donor transition multipole moments, as well as to signify the approximation of transition multipoles by physical multipoles. \autoref{Eq:Mfinal} facilitates the study of resonance energy transfer in inhomogeneous, absorbing, and dispersive media    
    for the cases where the sizes of the donor and the acceptor are comparable to the distance between them. Utilizing \autoref{Eq:Mfinal} the generalized form of the RET rate is calculated using the following expression, 
    \begin{eqnarray}
  \begin{aligned}
W^{Cas}_{\tau}&= K_{E2P}
\Big |
    \mathcal{E}^{Total}(\textbf{r}_{A_{1}},\textbf{r}_{D},\omega_{\alpha})\Big| ^{2}
   \quad \Big | \mathcal{E}^{Total}(\textbf{r}_{A_{2}},\textbf{r}_{D},\omega_{\beta})
   \Big| ^{2}\delta(E_{\delta})
    \label{RateE2P_Rs}
  \end{aligned}
  \end{eqnarray}
 The proposed theory provides a framework for simulation that has
significant computational advantages compared to calculating
the coupling factor utilizing dyadic Green’s functions.

\end{document}